# *Dirac reflection for a Single-layer Graphene Quantum Well*

## A. Mhamdi[a], E. Ben Salem[a], S. Jaziri[b]

[a] *Laboratoire de Physique de la Matière Condensée, Faculté des Sciences de Tunis, Tunisia*
[b] *Département de Physique, Faculté des Sciences de Bizerte, Tunisia*

We address the problem of Dirac fermions' graphene quantum well and we focus on the low energy approximation for the Hamiltonian of the system where the former can be described by a Dirac-like Hamiltonian. Interesting relations are obtained and used to discuss the influence of the spin-orbit coupling, which induces an effective mass-like term $mv_F^2$, on the transport properties of Single-layer graphene quantum well. It's found that the reflection probability of incident electrons is sensitive to the effective mass-like term $mv_F^2$. This can be explained by the dependence of *R* on the incident electrons' direction and their energies. Notably, we found that the reflection probability for massive fermions with a very small angle, i.e. the wave-vector along the transport direction is zero in the GQW, can be greatly suppressed.

Graphene is the first truly two-dimensional condensed matter system discovered experimentally in the group of A. Geim at the University of Manchester in 2004. Ever since graphene was first successfully produced experimentally, plenty of intriguing properties from its strictly 2D structure and mass-less Dirac behavior of low-energy excitation have been intensively investigated. Its peculiar properties do not only arouse pure scientific curiosity but also suggest possible practical applications.

As a single sheet of carbon, the physics of monolayer graphene devices has attracted a great deal of attention [1]. In fact, it's a honeycomb network where strong in-plane $sp^2$ hybrid bonds guarantee chemical stability and the interference of the remaining $p_z$ electrons in the honeycomb is responsible for physical properties. Accordingly, the exploration of monolayer graphene remains one of the most animated areas of research in condensed matter physics [2, 3, and 4].

Graphene and its few layer cousins constitute a unique electron system, and double-gated graphene devices have been a particularly attractive platform for investigation of electric properties in zero magnetic field. In other words, the finite deep confining potential serves well as a plat form for developing the physics that governs the dimensional systems and particularly for describing electrons bound to a thin sheet of conducting material [5]. Therefore, many studies were concerned about revealing the electronic properties of the finite quantum well based on a single-layer of graphene which is often realized with a thin layer of a graphene medium, embedded between other semiconductor layers of wider band-gap. However, to our knowledge, in the list of graphene's many remarkable properties, reflection probability of electrons propagating along the quantum well is notably absent. If it were possible to find a way to induce the former, it could improve the performance and enable more efficient integration of a variety of promising device concepts including nanoscales quantum interference devices. From a theoretical perspective, it is interesting to investigate the reflection probability of Dirac particles through the electrostatic monolayer graphene quantum well of constant depth and thickness. To this end, we explore the electronic reflection by analytical and numerical calculations.

The Fermi level in a neutral graphene sheet (a monolayer of carbon atoms with hexagonal lattice structure) is pinned near the corners of the hexagonal Brillouin zone which determine two non-equivalent valleys ***K*** and ***K'***. Actually, graphene layer consists of two triangular sublattices. The low-energy band is gapless and electronic states are found near two cones. When graphene is not doped, its Fermi level passes through the cone apexes. In such a situation, if one is interested in the low energy description, only the states near the cones must be accounted and the behavior charge carriers (electrons and holes) in each of the two valleys is governed by the 2D Dirac Hamiltonian [6,7].

Although in graphene the confinement by electrostatic potentials seems to be a difficult job by virtue of the absence of backscattering, which is a direct consequence of the pseudo chiral behavior of Dirac particles, theoretical [8] and experimental [9] works have proposed various schemes to study and overcome this "difficulty": single-layer strips, gated nanoribbons, gated and/or doped bilayers, etc.

Our analysis was inspired by an insightful paper of Peeters [8] who demonstrated a constructive interference between confined states and unbound states at the edge of the free particle by studying the resonance transmission of electrons across a graphene quantum well GQW with energies above the confining walls. Note that the model system we are interested in has been analyzed in [8]. However, in the present work we study other features of such a system to deal with different issues which concern the reflection probability for Dirac fermions. It's obvious that that region I and III are similar but different with respect to region II. For the sake of simplicity, in this study, we neglect the microscopic details of the interaction effects, such as the inter-valley coupling and the spin–orbit interaction. The sample width *w* is assumed to be so large that the edge effects can be neglected [10].

On this wise, the model system we are interested in is a graphene quantum well system. A schematic plot of the potential profile of this quantum well is given in figure 1. According to this configuration, we decompose the present system into three regions.

Let us first consider a square quantum well of depth $U_0 \geq 0$ and width *L* on which an electron of energy ($E \geq U_0$) is incident.

As the band structure of graphene has two valleys, which are decoupled in the case of a smooth edge, we focus on a single **K** point (valley). In fact, in graphene we can realize potential steps that are smooth ($L \gg a$), on the lattice scale $a = 1,42 \text{ Å}$ and therefore do not induce inter-valley scattering as the distance between the valleys in reciprocal space is $|\mathbf{K} - \mathbf{K}'| \sim 1/a$ [11].

Hence, for such potentials: see Fig. 1, valleys are decoupled and electrons, in the continuum limit, can be described by a single valley *2D Dirac Hamiltonian* including a diagonal effective mass-like term $mv_F^2$:

$$[v_F(\vec{\sigma} \cdot \vec{p}) + U + mv_F^2 \sigma_z] \Psi = E \Psi \quad (1)$$

Where the 'isospin' Pauli matrices $\sigma = (\sigma_x, \sigma_y)$ operate in the space of the charge carriers amplitude on two sites (A and B) in the unit cell of a hexagonal crystal, $\vec{p} = (p_x, p_y)$ is the momentum operator, and $v_F = 10^6 m \, s^{-1}$ is the Fermi velocity.

Here $U(x)$ is the one dimensional potential, *E* is the state energy and the term proportional to $mv_F^2$ introduces energy gap which may represent e.g. the effect of spin-orbit coupling or the interactions with the substrate.

In a given valley, since there are two atoms in graphene's unit cell, it is convenient to describe the single-electron wave of graphene as a two-component spinor [1, 12, and 13], $\Psi = [\Psi_A, \Psi_B]^T$. *T* is denoting the transpose of the row vector. The two components of *Ψ* refer to the two sublattices in the two-dimensional honeycomb lattice of carbon atoms (A and B).

In the presence of a one dimensional confining potential and due to the translation invariance in the *y*-direction, we can parameterize solutions $\Psi_A(x, y) = \phi_A(x)e^{ik_y y}$ and $\Psi_B(x, y) = i \phi_B(x)e^{ik_y y}$ by the conserved longitudinal momentum.

And Eq. (1) can be written as:
$$\begin{pmatrix} mv_F^2 & v_F(p_x - ip_y) \\ v_F(p_x + ip_y) & -mv_F^2 \end{pmatrix} \begin{pmatrix} \Psi_A \\ \Psi_B \end{pmatrix} = (E - U) \begin{pmatrix} \Psi_A \\ \Psi_B \end{pmatrix}$$

This can be presented in the following dimensionless form:

$$\begin{cases} \frac{d\phi_B(\xi)}{d\xi} + \beta \phi_B(\xi) = (\varepsilon - u - \Delta)\phi_A(\xi) & (2) \\ \frac{d\phi_A(\xi)}{d\xi} - \beta \phi_A(\xi) = -(\varepsilon - u + \Delta)\phi_B(\xi) & (3) \end{cases}$$

Where: $\xi = x/L$, *L* is the characteristic spatial scale of the potential variation, $\beta = k_y L$, $\varepsilon = \frac{EL}{\hbar v_F}$, $u = \frac{UL}{\hbar v_F}$ and $\Delta = \frac{mv_F^2 L}{\hbar v_F}$. These equations can be decoupled and formally written as a differential equation of second order for $\phi_A$:

$$\frac{d^2\phi_A(\xi)}{d\xi^2} + \frac{du/d\xi}{\varepsilon - u + \Delta} \frac{d\phi_A(\xi)}{d\xi} + \left[ (\varepsilon - u)^2 - \beta^2 - \beta \frac{du/d\xi}{(\varepsilon - u + \Delta)} - \Delta^2 \right] \phi_A(\xi) = 0 \quad (4)$$

$du/d\xi$ is the derivative of the potential. For a square well, the analytical solution is obtainable for constant $U(x)$ and Eq. (4) becomes then:

$$\frac{d^2\phi_A(\xi)}{d\xi^2} + f(k_y)\phi_A(\xi) = 0 \quad (5)$$

Where: $f(k_y) = [(\varepsilon - u)^2 - \beta^2 - \Delta^2]$.

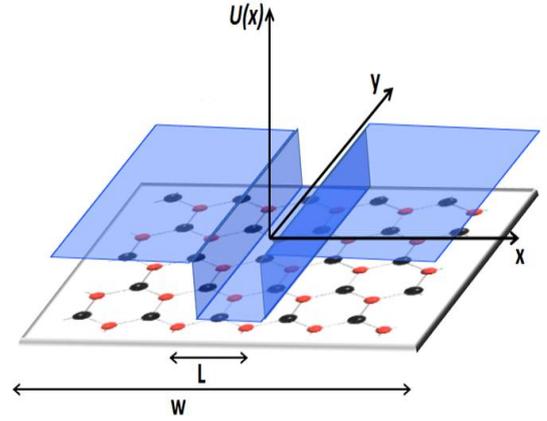

**Fig. 1:** A squared quantum of width *L* well on a graphene monolayer of width *w*.

Notice that $\phi_A(\xi)$ depends on $k_y$ through $\beta = k_y L$. And a similar expression holds for $\phi_B(\xi)$. This allows to write the solution of the spinor components in the form: $\phi_A(x) = \phi_A(x, k_y)$ and $\phi_B(x) = \phi_B(x, k_y)$.

The equation **(5)** hasn't the form of the usual Schrödinger equation with potential *u* and eigenvalue ε. The behavior of the solution will depend on the value of $\beta$ which determines the sign of the quantity $f(k_y)$. These solutions may occur tunneling waves, standing waves for $f(k_y) < 0$ or they may arise a traveling wave's distribution for $f(k_y) > 0$.

Focusing, in this stage, on the standing waves (where $f(k_y) < 0$) which for massless fermions arise only from finite values $\beta$ above an energy-dependent cutoff and decay exponentially in the barrier regions. These waves describe electrons states confined across the well and propagating along it. In [8] it has been shown, via an analytical framework that it is possible to trap Dirac electrons in a graphene quantum well of finite width

$L=200\ nm$ and depth $U_0 = 50\ meV$. In such a case, the confinement can be produced either by etching or simply by the application of gate potentials.

At low energy, the spectrum of the bound states generated by the attractive character of the potential $U(x)$ is obtained from the solution of the transcendental equation [8] and drawn in Fig.2 for $u = 18,56$, and $\Delta= 0$.

One would expect that the bound states would be localized in the GQW, as they are in the non-relativistic quantum well where the states are quantized in the quantum well and delimited by $u = 18,56$ (that corresponds to $U_0 = 50\ meV$).

In Fig. 2, the quantized electron branches are plotted in red and show a remarkable dependence of the eigenvalues on the component of the wave vector with a cutoff at low wave vectors. Note that for large wave vectors, the eigenvalues aren't equally spaced. In Fig. 2 there are two different regions. The green region corresponds to the waves that can propagate along the quantum well. This region is delimited by the curve $\varepsilon = \sqrt{(k_yL)^2 + \Delta^2} + u$ which represents the limit of the free-electrons continua. A fundamental difference with the corresponding results for Schrödinger-type electrons is that the spectrum of energy depends on $k_y$ instead of the wave vector $k_x$. A prenominal difference with non-relativistic electron states is the existence of a mixing between free and bound states at the former limit. Then for the blue region, it is delimited by $\varepsilon = -\sqrt{(k_yL)^2 + \Delta^2} + u$ and corresponds to free-holes that propagate in the system by means of the *Klein tunneling* mechanism.

At low wavevectors, there is an immediate conversion of confined electrons to free holes resulting from the absence of a gap in the spectrum and from the chiral nature of the quasi-particles in graphene.

We stress the particularity that the quantized energy levels of bound states exceed the limit $\varepsilon = 18,56$ which is in striking contrast to the case of non-relativistic QW. One notices the existence of electronic bound states or above quantum well confined states in the above part of the quantum well. One can notice also that there is evidence of the existence of quantized electron branches that intercept the free particles regions. Accordingly, there is a constructive interference between confined states and unbound electron states that are resonantly transmitted across the quantum well for $\varepsilon = \sqrt{(k_yL)^2 + \Delta^2} + u$. Fig. 3 shows clearly that the confined states are not allowed for $k_y = 0$.

An intriguing behavior is found above the well depth for $\varepsilon > \sqrt{(k_yL)^2 + \Delta^2} + u$ (this corresponds to the green region). In order to study this behavior, we consider the scattering states (where $f(k_y) > 0$) which describe free electrons, free holes, and mixed states. And we assume energies to be greater than $\sqrt{(k_yL)^2 + \Delta^2} + u$.

Explicit analytical expressions of the wave function are essential for demonstrating the standard properties and the peculiarities of quantum phenomena.

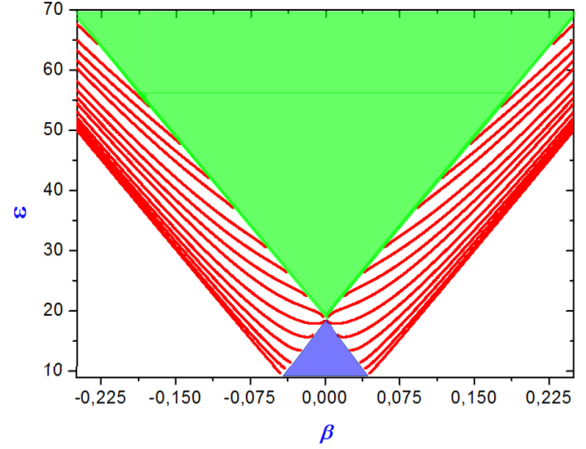

**Fig. 2:** Spectrum of confined states in graphene quantum well: ε versus β (for color region see in the text).

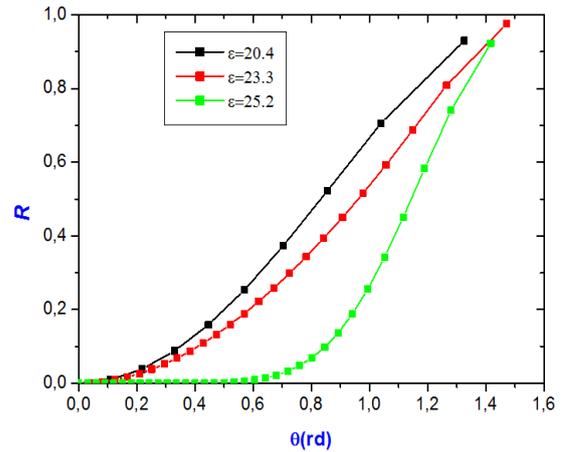

**Fig. 3:** Reflection $R$ versus the angle of incidence for different dimensionless energies.

So, for $f(k_y) > 0$, the propagating solutions at a given incident energy ε, are represented by:

$$\phi_A(\xi) = \begin{cases} e^{i\alpha\xi} + r\ e^{-i\alpha\xi} & -\dfrac{w}{2L} < \xi < -\dfrac{1}{2} \\ A_2\ e^{ik\xi} + B_2 e^{-ik\xi} & |\xi| \leq \dfrac{1}{2} \\ t\ e^{i\alpha\xi} & \dfrac{1}{2} < \xi < \dfrac{w}{2L} \end{cases} \quad (6)$$

Where $\alpha = ((\varepsilon - u_0)^2 - \beta^2 - \Delta^2)^{1/2}$ and
$k = (\varepsilon^2 - \beta^2 - \Delta^2)^{1/2}$.

The reflection and transmission amplitudes across the graphene quantum well are denoted by *r* and *t*, respectively.

From Eq. **(3)** $\phi_B$ can be given by:

$$\phi_B(\xi) = \begin{cases} g_- e^{i\alpha\xi} + r\, g_+ e^{-i\alpha\xi} & -\dfrac{w}{2L} < \xi < -\dfrac{1}{2} \\ A_2 f_- e^{ik\xi} + B_2 f_+ e^{-ik\xi} & |\xi| \leq \dfrac{1}{2} \quad (7) \\ t g_- e^{i\alpha\xi} & \dfrac{1}{2} < \xi < \dfrac{w}{2L} \end{cases}$$

Where $g_\pm = \dfrac{\beta \pm i\alpha}{\varepsilon - u_0 + \Delta}$ and $f_\pm = \dfrac{\beta \pm ik}{\varepsilon + \Delta}$.

The amplitude of the sinusoidal waves gives the strength of the probability density current and can then be determined from the experimental flux of particles. The amplitude of the wave $e^{i\alpha\xi}$ for $\xi < -1/2$ is determined by the flux of particles from a source at $\xi = \dfrac{-w}{2L}$. These particles can be reflected or transmitted by the potential leading to a reflected wave $r\, e^{-i\alpha\xi}$ in the region $\xi < -1/2$ and a transmitted wave $t\, e^{i\alpha\xi}$ in the region $\xi > 1/2$. This is a 1D scattering problem. We content ourselves with noting that all energies for $E > U_0$ are possible. In this energy regime, we have a continuum of allowed energies.

In order to have solutions with sensible probability densities, both the probability density and the probability density currents must be continuous functions of $\xi$. These solutions can be ensured by requiring the continuity of only $\Psi(\xi)$ and not of its derivative $\Psi'(\xi)$. For the Schrödinger equation, we would have had to use the continuity of the wave function and that of its derivative as well. Here the two-component spinor allows the same number of equations just from the continuity of the wave function. Therefore, let us drive the adequate conditions of $\phi_A$ and $\phi_B$ at the borders of the QW, for $\xi = \pm\dfrac{1}{2}$, which lead to the following system:

$$\begin{cases} e^{-i\frac{\alpha}{2}} + r e^{i\frac{\alpha}{2}} = A_2 e^{-i\frac{k}{2}} + B_2 e^{i\frac{k}{2}} \\ A_2 e^{i\frac{k}{2}} + B_2 e^{-i\frac{k}{2}} = t e^{i\frac{\alpha}{2}} \\ g_- e^{-i\frac{\alpha}{2}} + r g_+ e^{i\frac{\alpha}{2}} = A_2 f_- e^{-i\frac{k}{2}} + B_2 f_+ e^{i\frac{k}{2}} \\ A_2 f_- e^{i\frac{k}{2}} + B_2 f_+ e^{-i\frac{k}{2}} = t g_- e^{i\frac{\alpha}{2}} \end{cases}$$

Thus, we obtain a set of linear equations for the parameters $A_2, B_2, r$ and $t$. By matching up these coefficients we find the following expressions for $r$ and $t$:

$$r = \dfrac{2\sin(k) e^{-i\alpha} e^{ik}(g_- f_+ - f_- f_+ - g_-^2 + f_- g_-)}{(g_+ - f_+)(g_- - f_-) e^{2ik} - (g_+ - f_-)(g_- - f_+)} \quad (8)$$

$$t = \dfrac{(g_+ - g_-)(f_+ - f_-) e^{-i\alpha} e^{ik}}{(g_+ - f_+)(g_- - f_-) e^{2ik} - (g_+ - f_-)(g_- - f_+)} \quad (9)$$

These results will be used to deal with the reflection and transmission coefficients and discuss their different features as a function of the well depth, the incident energy and the angle of incidence.

In fact, the transmission coefficient $T$ measures the proportion of the transmitted particles. But since the wave emerging forward involves both the intensity of the scattered wave and an interference term. It is, therefore, impossible to distinguish forwardly between the particles which are really released by the well and those that would come to without being interacted with the potential. Nevertheless, the coefficient $R$ measures the importance of the back-scattering which is caused by the action of the potential well on the charge carriers. This finally allows us to admit that the reflection coefficient is the coefficient characterizing the efficiency of scattering by the quantum well potential.

Now, we wish to obtain the total reflection and transmission probabilities $R$ and $T$, respectively, which are obtained by taking the square of the modulus of the reflection amplitude $r$ and the transmission amplitude $t$ given by (8) and (9) as $R = |r|^2$ and $T = |t|^2$.

$R$ and $T$ can then be written as:

$$R = \dfrac{4\sin^2(k)(r_1^2 + r_2^2)}{\{r_3 \cos(2k) - r_4\}^2 + \{r_5 \sin(2k)\}^2} \quad (10)$$

Where: $r_1 = 2\beta^2(\varepsilon - u_0 + \Delta)(\varepsilon + \Delta) - (\beta^2 + k^2)(\varepsilon - u_0 + \Delta)^2 - (\beta^2 - \alpha^2)(\varepsilon + \Delta)^2$ and

$r_2 = 2\alpha\beta(\varepsilon + \Delta)^2 - 2\alpha\beta(\varepsilon - u_0 + \Delta)(\varepsilon + \Delta)$, $r_3 = \beta^2 u_0^2 + (\alpha\varepsilon + \alpha\Delta - k\varepsilon + ku_0 - k\Delta)^2$, $r_4 = \beta^2 u_0^2 + (\alpha\varepsilon + \alpha\Delta + k\varepsilon - ku_0 + k\Delta)^2$ and $r_5 = \beta^2 u_0^2 + (\alpha\varepsilon + \alpha\Delta - k\varepsilon + ku_0 - k\Delta)^2$.

$$T = \dfrac{16\,\alpha^2 k^2 (\varepsilon - u_0 + \Delta)^2 (\varepsilon + \Delta)^2}{\{t_1 \cos(2k) - t_2\}^2 + \{t_3 \sin(2k)\}^2} \quad (11)$$

Where: $t_1 = \beta^2 u_0^2 + (\alpha\varepsilon + \alpha\Delta - k\varepsilon + ku_0 - k\Delta)^2$, $t_2 = \beta^2 u_0^2 + (\alpha\varepsilon + \alpha\Delta + k\varepsilon - ku_0 + k\Delta)^2$, and $t_3 = \beta^2 u_0^2 + (\alpha\varepsilon + \alpha\Delta - k\varepsilon + ku_0 - k\Delta)^2$.

The equation (9) of the reflection coefficient $R$ can be simplified to this analytical solution where $\Delta \neq 0$:

$$R_{\Delta \neq 0} = \dfrac{u_0^2 \left(\dfrac{\sqrt{\varpi_3}}{\alpha k L(\varepsilon + \Delta)(u_0 - \varepsilon - \Delta)}\right)^2 \sin^2(kL)}{1 + \left(\dfrac{2\sqrt{\varpi_1 \varpi_2}}{\alpha k L(\varepsilon + \Delta)(u_0 - \varepsilon - \Delta)}\right)^2 \sin^2(kL)} \quad (12)$$

Where:

$k = \sqrt{(\varepsilon^2 - \beta^2 - \Delta^2)/L^2}$, $\varpi_1 = [\beta^2 u_0^2 + (\alpha\varepsilon + \alpha\Delta - kL\varepsilon + kLu_0 - kL\Delta)^2]$,

$\varpi_2 = [\beta^2 u_0^2 + (\alpha\varepsilon + \alpha\Delta + kL\varepsilon - kLu_0 + kL\Delta)^2]$ and

$\varpi_3 = \alpha^2 \beta^2 (\varepsilon + \Delta)^2 + [-\Delta^3 + \Delta^2(u_0 - 2\varepsilon) + \Delta(\beta^2 - \varepsilon^2 + 2\varepsilon u_0) - \varepsilon\beta^2]^2$.

Now, we move on to the case where Δ= 0 and the coefficient R takes this form:

$$R_{\Delta=0} = \frac{u_0^2 \left(\frac{\beta\sqrt{\beta^2+\alpha^2}}{\alpha kL(u_0-\varepsilon)}\right)^2 \sin^2(kL)}{1+\left(\frac{\sqrt{X_1 X_2}}{2\alpha kL\varepsilon(u_0-\varepsilon)}\right)^2 \sin^2(kL)} \quad (13)$$

Here:

$k = \sqrt{(\varepsilon^2-\beta^2)/L^2}$, $X_1 = [\beta^2 u_0^2 + (\alpha\varepsilon - kL\varepsilon + u_0)^2]$ and

$X_2 = [\beta^2 u_0^2 + (\alpha\varepsilon + kL\varepsilon - kLu_0)^2]$.

Regardless of the incident energy chosen, if the potential is zero (or very small enough) the charge carriers will obviously pass down. As comparison to the Schrödinger case, this result is in accordance with the corresponding non-relativistic problem. As a matter of fact, the reflection coefficient measures the importance of the back-scattering which is caused by the action of the potential well on the charge carriers. We note immediately that for $u_0 = 0\ meV$, $u_0$ is giving by $\frac{U_0 L}{\hbar v_F}$, the coefficient R is zero.

On the one hand, the analytical calculations, show that the reflection through the quantum well can be calculated even for energy greater than $U_0$. On the other hand, the reflection coefficient (and the transmission coefficient as well) depends on the direction of propagation. In other words, it depends on the incident angle θ of the incident electrons (see below). In contrast to the non-relativistic case, where the problem does not depend on β and thereafter on $k_y$. We note that the electron has a free-like propagation along the y direction, which means that each value of $k_y$ imposes its own reflection coefficient.

We will see below that the behavior of Dirac particles is very different. To elucidate further the difference between quantum transport of Dirac fermions and non-relativistic 2D electrons, it is interesting to explore the reflection probability as a function of the angle of incidence θ. Such that, we will plot the coefficient R versus the angle θ which is giving by the following equation:

$$\theta = \arctan\left(\frac{k_y L}{\alpha(k_y)}\right) \quad (14)$$

The reflection coefficient R for a potential of the depth $u = 18,56$, a width L=200 nm and for Δ = 0 is depicted for different dimensionless electron energies as indicated in Fig. 3.

The present situation is quite different from the conventional one. In Fig. 3 one notices that the reflection depends strongly on the direction of the incident electron waves, for θ = 0, the coefficient R is always equal to zero ($R(\theta = 0) = 0$) which is a signature of the *Klein Tunneling* [14]. This comes as a result of the absence of *back-scattering*. In fact, the most striking feature in the reflection probability is that for a normal incidence where $k_y = 0$ and for any dimensionless energy the effect of the potential turns out less important and we obtain imperfect reflection. Therefore, normally-incident electrons are always perfectly transmitted.

For the sake of completeness, one has to resort to R's evolution with incident electron energies. In this section, we move on to the presentation of R versus the incident energy for fixed angles θ as shown in Fig. 4.

Working out the problem for different angles of incidence allows us to learn more how to get a perfect reflection and how to position the incident beam of electrons. Therefore, it's worth to represent R for different incidence angles.

We show, in Fig. 4, the variation of R with incident energy for a potential of depth $U_0 = 50\ meV$, a width L=200 nm and for $mv_F^2 = 0$. The variation of R with energy shows characteristic oscillations with maxima and zeros and we can distinguish two types of specific physical situations.

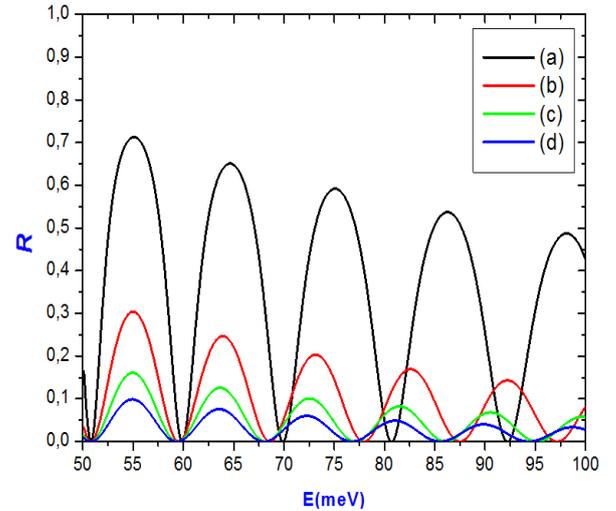

**Fig.4:** Dependence of the reflection coefficient on the incident energy with different incident angles: (a) θ=π/3, (b) θ=π/5, (c) θ=π/7, and (d) θ=π/9 for Δ= 0.

The oscillatory behavior of R is due to the sine in Eq. (13) and due to the effect of resonance which could be interpreted as a quasi-stable intermediate state. If we consider this situation, the maxima could have a significant effect. In fact, the maxima of R roughly correspond to the values of the energy that satisfy the following relation $kL = (n + \frac{1}{2})\pi$ and the potential acts maximally on the particles. Under such conditions, the particles are somehow "trapped" in the well; so that they pass into the interaction region a maximum time before getting out.

We notice also that the minima, corresponding to the transparency of the system, always reach zero for $kL$ equal to an integer multiple of $\pi$ ($kL = n\pi$). In other words, there are particular values of the energy where the potential does not "back-scatter" the charge carriers. Hence, electrons may undergo a significant back-scattering and could pass down all through the interaction region, as if the well was perfectly transparent. Such a situation reminds us the Ramsauer-Townsend effect.

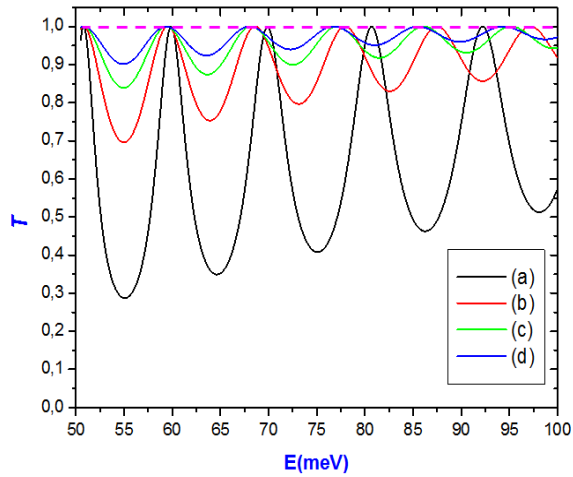

**Fig. 5**: The transmission coefficient as a function of incident energy with different incident angles: (a) θ=π/3, (b) θ=π/5, (c) θ=π/7, and (d) θ=π/9.

The relativistic reflection coefficient displays aperiodic minima as a function of the incident energy. Moving away from $E = 50\ meV$, the period (equal to $\frac{2\pi}{kL}$) increases up significantly. While the locations of the minima shift toward the low energy with decreasing of the incident angle. And we find that the reflection probability of electrons with a very small incident angle (>π/9), i.e. the wave-vector along the transport direction is zero, can be greatly suppressed.

The intensity of the maxima decreases gradually as one moves away from $E = 50\ meV$ and decreases the angle of incidence. This shows that the maxima are particularly more important as the energy E is close to the value of the depth of the well $U_0$.

Now, let us look to what happens with $T$ for different angles of incidence $\theta$, for L=200 nm, $U_0 = 50\ meV$ and $mv_F^2 = 0\ meV$.

In Fig. 5, $T$ versus E, an oscillatory feature is observed. One clearly remarks that the intensity of the minima of the coefficient $T$ decreases when decreasing the angles of incidence. The most striking feature in the transmission probability is that at $kL$ being equal to an integer multiple of $\pi$, we obtain perfect transmission ($T = 1$) for any angle of incidence and any depth of the well. The dashed curve presents $T$ added to $R$.

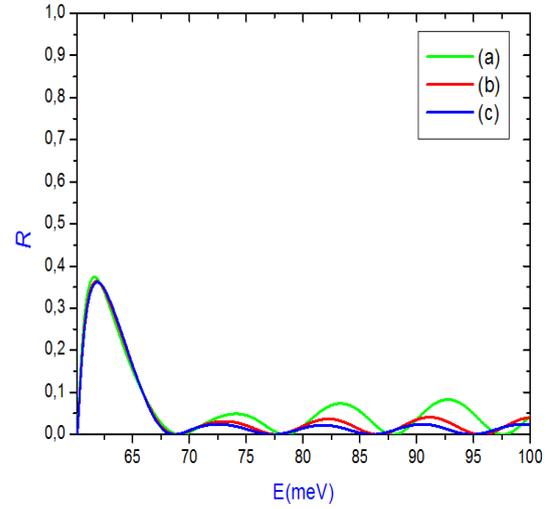

**Fig. 6:** The reflection coefficient as a function of incident energy with different incident angles: (a) θ=π/5, (b) θ=π/7, and (c) θ=π/9 for $\Delta \neq 0$.

Our next step is to examine the influence of the effective mass-like term $mv_F^2$ on the reflection coefficient which is presented in Fig.7 for a potential of the depth $U_0 = 50\ meV$, a width $L=200\ nm$ and for $mv_F^2 = 10\ meV$.

Fig.6 shows the effect of nonzero effective mass. An oscillatory behavior is observed in this case as well. The introducing of the mass term makes the intensity of the maxima of the reflection coefficient more sensitive. It is found that when the incident angle is very small (>π/9) the reflection probability has been significantly reduced and it is approximately equal to zero ($R \approx 0$) for energy high than 67,5 meV. Such that, one can see that the transmission probability is robust to the mass-like effect.

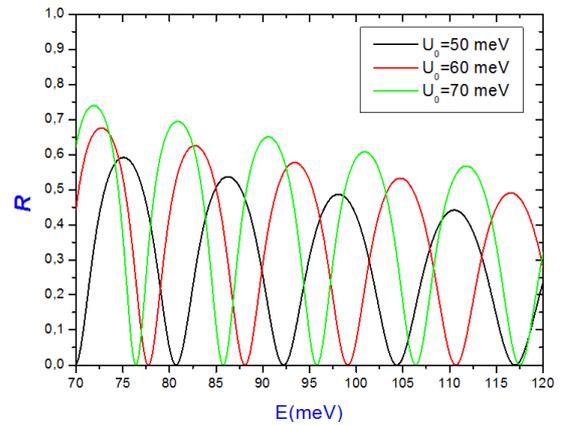

**Fig. 7:** The reflection coefficient as a function of incident energy with different depth $U_0$ for θ=π/3.

It will be instructive to examine the effect of the depth of the quantum well on the reflection coefficient, for this end let's plot $R$ versus the energy for different $U_0$ for L=200 nm and $mv_F^2 = 0\ meV$ (see Fig.7).

In Fig. 7, it's clear that the minima positions are related to the energies of the corresponding depth of the QW. These locations shift toward the low energy with increasing the depth of the well. As the maxima, their intensities become important due to the stronger confinement effect at large incident angle.

We believe that our study is crucial for a better understanding of the electronic properties of Dirac particles for a quantum well consisting of a single-layer graphene. We have evaluated the reflection properties of a monolayer QW in the low energy approximation using the Dirac-like Hamiltonian including the effective mass-like term $mv_F^2$. The reflection probability $R$ of electrons is characterized by the following features: We have found that $R$ exhibits a strong dependence on the direction of the incident electron wave vector. We have demonstrated that a 2D Dirac electron normally incident can't be backscattered and we obtained imperfect reflection. A further aspect of our results is that $R$ with a very small angle of incidence, i.e. the wave-vector along the transport direction is zero, can be greatly suppressed. Accordingly, the transmission probability is robust to the effective mass-like correction.